\documentclass[%
 aip,
 amsmath,amssymb,
 reprint,%
]{revtex4-1}
\makeatletter
\def\@email#1#2{%
 \endgroup
 \patchcmd{\titleblock@produce}
  {\frontmatter@RRAPformat}
  {\frontmatter@RRAPformat{\produce@RRAP{*#1\href{mailto:#2}{#2}}}\frontmatter@RRAPformat}
  {}{}
}%
\makeatother

\usepackage{graphicx} 
\usepackage{xcolor}
\usepackage{subcaption} 
\usepackage{amsmath} 
\usepackage{amssymb} 
\usepackage{hyperref}

\usepackage{amsthm}

\newtheoremstyle{boldtitle}
  {3pt}    
  {3pt}    
  {\itshape} 
  {}       
  {\bfseries} 
  {:}      
  {1em}    
  {}       
\theoremstyle{boldtitle}
\newtheorem{definition}{Definition}

\begin{document}

\preprint{AIP/123-QED}

\title{Constructing directed networks with a desired minimum balanced coloring}
\author{Jonathan Martinez}
\affiliation{Department of Mechanical Engineering, University of New Mexico, Albuquerque, NM, 87131}
\author{Teresa Radice}
\affiliation{Dipartimento di Matematica e Applicazioni ``R. Caccioppoli'', Universit\`a Federico II, Napoli, Italy, 80126}
\author{Francesco Sorrentino}
\thanks{Corresponding author: \href{mailto:fsorrent@unm.edu}{fsorrent@unm.edu}}
\affiliation{Department of Mechanical Engineering, University of New Mexico, Albuquerque, NM, 87131}

\date{April 2025}

\begin{abstract}
Biological, social, and technological networks are made up of many interacting nodes coupled by directed connections. Moreover, the study of empirical networks has shown that often groups of their nodes are `symmetric' to one another, i.e., they are structurally identical to one another within such networks. 
Many of the popular network-generating algorithms do not control for the emergence of symmetries.
Here, we present an algorithm focused on directed networks that enables the user to generate what are called `expanded' directed networks, which transform each node of a base network into a cluster with a desired number of nodes, after checking for the expansion to be feasible. Analytical conditions are provided for an expansion to be feasible and for the generation of the expansion or minimal expansion, i.e., a feasible expansion with the least number of nodes. The algorithm is used to construct synthetic networks with prescribed symmetries.
\end{abstract}

\maketitle

\textbf{This work introduces a systematic algorithm for generating directed networks with prescribed symmetries by constructing expansions from a given quotient network. The method enables researchers to synthesize realistic network models with controllable symmetry structure, facilitating studies of symmetry-driven dynamics such as cluster synchronization in biological, social, and technological systems.}

\section{Introduction}
Networks are a commonplace occurrence in a broad range of fields, including biology, physics, engineering, and the social sciences. 
A common feature of several empirical networks is the presence of symmetries within subsets of the network nodes \cite{makse2025symmetries}. For the moment, we intentionally use the word `symmetry' in a loose sense, with precise definitions introduced in what follows.
One of the advantages of this approach is that it allows one to derive a simplified description of a network into a less complex representation, dubbed the base or `quotient network'. A large literature has investigated the relation between these symmetries and the network dynamics, such as in the case of cluster synchronization \cite{NC,SA,blaha2016symmetry,klickstein2019symmetry,gambuzza2020controlling,lodi2020analyzing,sorrentino2020group}.

In physics, symmetries are defined in terms of group theory. However, the most common occurrence of symmetry in biological networks is in the form of `fibration symmetries' \cite{grothendieck1959technique}, following the definition originally provided by Grothendieck in algebraic geometry.

An important distinction in graph theory is between undirected and directed networks. This paper draws its motivation from the observation that (i) most real networks are directed and (ii) most real networks have symmetries. 
Within the large literature that studies networks  and their symmetries, an important issue is how to generate networks with prescribed symmetries. This has previously been studied by Klickstein et al. \cite{klickstein2018generating, klickstein2018generating2} for undirected networks. However, the large majority of real networks are directed, hence it becomes important to develop methods to generate directed networks that possess specific symmetries. This is precisely the goal of this paper.

\section{{Preliminaries}}
Here we deal with simple directed networks, which are particular networks that have neither self-loops nor repeated edges in a given direction between any two vertices. In this paper,
these are defined by a 3-tuple \(\mathcal{G=(V,E},A\mathcal{)}\), where \(\mathcal{V}=\{1, ...,N\}\) is the set of the $N$ network vertices and $\mathcal{E} =\{e_1,e_2...,e_M\} \subseteq \mathcal{V\times V}$  is the set of $M$ network directed edges. We can write a directed edge \(e_\ell=(j \rightarrow i\)), \(i,j \neq i\in\mathcal{V}\), \(\ell=1,...,M\).  The set of edges \(\mathcal{E}\) is defined such that for any ordered pair \((j \rightarrow i)\in V\times V\), there is at most one edge from \(j\) to \(i\). 
An edge is graphically represented by an arrow so that the source/tail is $s(e_\ell)=j$ and the target/head is $t(e_\ell) = i$ \cite{harary1969graph,berge1973graphs}. In what follows, we will refer to the tail/source of a directed edge as the parent vertex and to the head/target of a directed edge as the child vertex.

The adjacency matrix \({A} \in \mathbb{R}^{N \times N}\) is defined such that \(A_{ij}=1\) if there is a directed edge \(e_\ell=(j \rightarrow i)\) going from node $j$ to node $i$
and \(A_{ij}=0\) otherwise \cite{makse2025symmetries}. We emphasize that we use the notation in which \(A_{i,j}\Rightarrow A_{child, parent}\). 
The variable \(k^{in}_i=\sum_j A_{ij}\) represents the in-degree of node \(i\), so that the number of edges whose target is \(i\) is \(k^{in}_i.\) \cite{boldi2002fibrations}. 

Next we provide the definitions of the input set of a node $i$ and of the input tree of a node $i$.

\begin{definition} \textbf{Input set of node $i$.}
Given a network \(\mathcal{G=(V,E,}A\mathcal{)}\), the input set of node \(i\in\mathcal{V}\) is \(I_i=\{i\}\cup\{e|e\in\mathcal{E},t(e)=i\}\), where \(t(e)\) is the  target node \(i\) of the edge. \cite{boldi2002fibrations}
\end{definition}

\begin{definition} \textbf{Input tree of node $i$.}
 If \(I_i={i}\), the input tree \(T_i\) is the target node itself. Alternatively, if \(I_i=\{i,e_1,...,e_{k_{in}}\}\), \(T_i\) is the input tree of the target node \(i\) that has \(k_{in}\) edges and nodes, subtrees to those \(k_{in}\) nodes, \(T_{s(e_1)},...,T_{s(e_{k_{in}})}\), subtrees to those, and so on. The node itself resides at level 1 of the tree, while its parents reside at level 2 of the tree. The subtrees proceed from level 3 to \(N\), where \(N\) is the number of nodes in the network \cite{morone2020fibration,aldis2008polynomial,stewart2007lattice,boldi2002fibrations}. 
\end{definition}

If a node is part of a cycle within the network, the input tree will continue forever. However, in practice, we can stop the calculation of the input trees at level $N-1$, following previous work \cite{angluin1980local,norris1995universal,GolubitskyStewartTorok2005}.

We say \(T_i\simeq T_j\) when two input trees \(T_i\) and \(T_j\) are isomorphic to each other \cite{harary1969graph,makse2025symmetries}. Because the isomorphism is an equivalence relation, it induces a partition of the set of the network nodes $\mathcal V$ into subsets that we call `clusters' or `fibers' (in this paper we will prefer the term clusters),  \(\mathcal{P}=\mathcal{\{C}_1,\mathcal{C}_2,...,\mathcal{C}_P\mathcal{\}}\), such that \(\cup^P_{k=1}\mathcal{C}_k=\mathcal{V}\) and \(\mathcal{C}_k\cap\mathcal{C}_l=\emptyset\) when \(k\neq l.\) When a node's input tree does not exhibit an isomorphism with any other input tree in a network, the node belongs to a cluster of its own. We will say that nodes that are in the same cluster are symmetric to one another. Fibration symmetries applied to \(\mathcal{G}\) are defined by isomorphisms between the input trees of a network.

\begin{definition} \textbf{Equitable partition of a network.}
A partition of a network into $P$ clusters \(\mathcal{\{C}_1,\mathcal{C}_2,...,\mathcal{C}_P\mathcal{\}}\) is said to be equitable only if each node in $\mathcal{C}_v$ receives the same number of edges $k_{v\mu}$ from each node in $\mathcal{C}_\mu$ where $v\neq\mu$ \cite{makse2025symmetries}.
\end{definition}

By their definition, the partition of a network using input tree isomorphisms is an equitable partition of the network. It can be shown that it is also the coarsest equitable partition, in the sense that it uses the least number of clusters $P$ \cite{makse2025symmetries}.
The coarsest equitable partition is unique for each network \cite{makse2025symmetries}.

Another important concept that we define in this paper is that of a coloring, a partition of the set of the network nodes $\mathcal{V}$ into subsets, where all the nodes in the same subset share the same color. 
 A balanced coloring partition is such that nodes with the same color \(\mu\) in subset \(\tilde{\mathcal{C}}_\mathcal{\mu}\) each receive the same number of edges \(k_{\mu v}\) from those in subset \(\tilde{\mathcal{C}}_v\), for all \(v\) and \(\mu\), which indicates an equitable partition. A minimum balanced coloring partition of the network is the above, but with the condition that it contains the least number of unique colors (or clusters) possible and, therefore, is the coarsest partition attainable for the network \cite{golubitsky2023dynamics}. There is a nice and efficient algorithm to compute the minimum balanced coloring of a graph, i.e., the \emph{color refinement algorithm}, originally proposed by Unger \cite{unger1964git} and further studied and developed in \cite{corneil1970efficient,cardon1982partitioning,belykh2011mesoscale}. The algorithm consists of three steps: (i) Initialization, start with the trivial partition, i.e., all nodes in one cluster; (ii) Refinement, for each cluster of nodes, check whether all nodes in it have the same number of directed edges from each other cluster; If not, split the cluster into sub-clusters accordingly; (iii) Termination, repeat until no further refinement is possible.
The algorithm stops at the minimum balanced coloring, i.e., the coarsest equitable partition for that network.
 
We have introduced two different partitions of the set of the network nodes $\mathcal{V}$, one which we have called the coarsest equitable partition and another which we have called the minimum balanced coloring. It can be shown that these two partitions coincide, i.e., they partition the set of the network nodes $\mathcal{V}$ into the same subsets ${\{C}_1,\mathcal{C}_2,...,\mathcal{C}_P\}$ \cite{makse2025symmetries}. In what follows we will simply refer to these subsets as clusters.

A quotient network is the most simplified representation of a network in which each cluster is collapsed into one node.   
Given a network \(\mathcal{G=(V,E,}A\mathcal{)}\) and an equitable partition of the set of nodes $\mathcal{V}$ into clusters $\mathcal{C}_1, \mathcal{C}_2,..., \mathcal{C}_P$, we can construct the non-simple directed quotient network \(\mathcal{Q=(V^Q,E^Q,}A^\mathcal{Q}\mathcal{)}\).

\begin{definition}\textbf{Quotient network.} 
We can formally define the compression of \(\mathcal{G}\) into its quotient network as \(\mathcal{Q}=\mathcal{G/P}\), where \(\mathcal{P}=\mathcal{\{C}_1,\mathcal{C}_2,...,\mathcal{C}_P\mathcal{\}}\) 
is a minimal equitable partition of the set of the network nodes (or equivalently a minimum balanced coloring.)
The quotient adjacency matrix \(A^\mathcal{Q}\)  $\in \mathbb{R}^{P\times P}$ is defined so that for each pair of clusters \(\mathcal{C}_k\) and \(\mathcal{C}_l\),
\begin{equation}\label{adj relation eq}
A^\mathcal{Q}_{kl}=\sum_{j\in\mathcal{C}_l}A_{ij}\quad\textrm{with}\quad i\in\mathcal{C}_k.
\end{equation}
\end{definition}

Note that the adjacency matrix of the quotient network \({A}^\mathcal{Q}=\{{A}^\mathcal{Q}_{ij}\}\) is not restricted to values of 0 or 1, unlike \({A}\), and can contain self-loops.   We emphasize that we adopt the notation according to which ${A}^\mathcal{Q}_{kl}>0$ indicates the presence of direct edge(s) going from node $l$ to node $k$ of the quotient network. 

\begin{figure}[h!]
    \centering
    \includegraphics[height=3.5in]{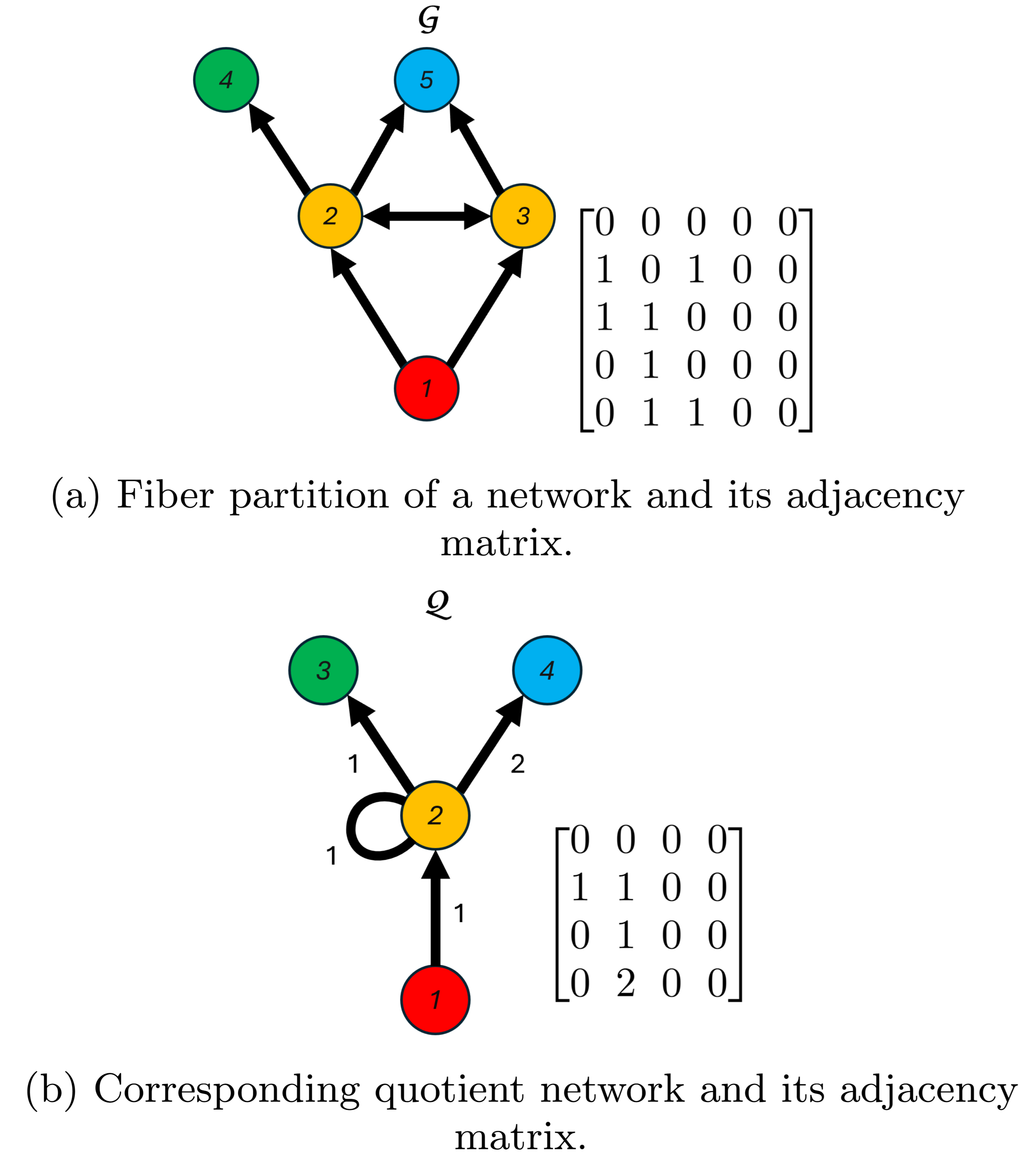}
    \caption{Example of a directed network and of the associated quotient network. Different colors are used to label nodes in different clusters.} \label{Intro Quotient Network}
\end{figure}

Figure \ref{Intro Quotient Network} shows an example of a network (\ref{Intro Quotient Network}a) and its corresponding quotient network (\ref{Intro Quotient Network}b) in which nodes that belong to the same cluster are shown with the same color. The clus ters we have defined associated with the quotient network are, by definition, a minimum balanced coloring \cite{makse2025symmetries}.
\\

\begin{figure*}[t]
    \centering
    \includegraphics[width=1\textwidth]{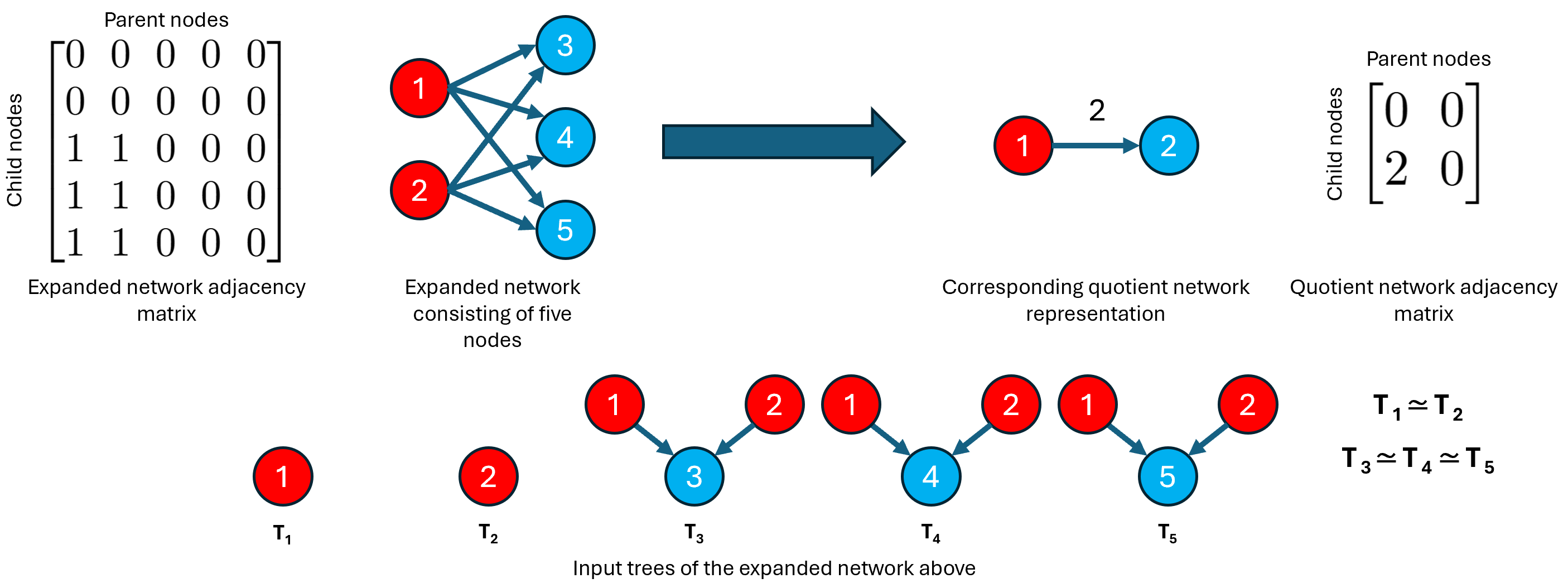}
    \caption{Example diagram of input tree representations of an expanded network which form clusters. The clusters enable the transformation of the expanded network into its quotient network.}
    \label{fig:Input trees}
\end{figure*}

Another example of a network with its associated quotient network is shown in Figure \ref{fig:Input trees}, where an original $N=5$-node network is transformed into the corresponding $P=2$-node quotient network. The adjacency matrices for both networks are shown as well. The input trees of all nodes in the expanded network are shown underneath the transformation. Clusters are indicated by color (red and blue). The cluster \(\mathcal{C}_1=\{1,2\}\) contains the red nodes,  given that trees \(T_1\) and \(T_2\) are isomorphic. Similarly, \(\mathcal{C}_2=\{3,4,5\}\). The expanded network is related to its quotient network according to the relation between \(A^Q \) and \(A\) given by Eq.\ \eqref{adj relation eq}.



\section{Problem Statement and Feasibility} \label{Methods}

In this section, we propose a method to generate an expanded network with a desired cluster partition from knowledge of two user-provided inputs: (i) the quotient network with $P$ nodes and (ii) a number of nodes that we want to place in each cluster of the expanded network $n_1,n_2,..,n_P$, where $n_i=|\mathcal{C}_i|$, $\sum_i n_i=N$. 

In order to generate an `expanded network' from the two inputs provided, namely the quotient network and a desired number of nodes for each cluster, we first need to address the question of feasibility, i.e., whether, for a given quotient network and desired number of nodes in each cluster, the expansion can be achieved.
If the expansion is feasible, we will then provide an algorithm that will output the expanded network(s). The algorithm is designed so that it generates a random network expansion among those that are possible. {It also forbids the generation of self-loops in the network expansions.} 

\begin{definition}{\textbf{Feasibility of a quotient network expansion.}} Given a quotient network $\mathcal{Q}$ with $P$ nodes, with adjacency matrix $A^\mathcal{Q}$ and a desired number of nodes for each cluster $n_1,n_2,...,n_P$,  there are two conditions that must be met for the expansion to be feasible: 
\begin{subequations}\label{Feasibility Condition Equations}
\begin{align}
    n_l\geq & A^\mathcal{Q}_{kl}  \quad \forall k \neq l\label{Outgoing Edge Condition}\\[5pt]
    n_l> & A^\mathcal{Q}_{ll}\label{Self-Loop Edges Condition}
\end{align}
\end{subequations}
\end{definition}

Constraint \eqref{Outgoing Edge Condition} indicates that the number of nodes in each cluster must be equal to or greater than the value of the highest weighted outgoing edge from that cluster. Note that for each cluster $l$, this condition requires checking the entries in the column $l$ of the matrix $A^\mathcal{Q}$ (except for entry $A^\mathcal{Q}_{ll}$.) 
Constraint \eqref{Self-Loop Edges Condition} indicates that for each cluster the number of nodes must exceed  the corresponding number of self-loops in $A^\mathcal{Q}$. The desired expansion is considered to be a minimal expansion when it contains the smallest number of nodes possible for every cluster of the quotient network. 
Therefore, a minimal feasible expansion is achieved when
\begin{equation} \label{minimimal}
n_l=\max{\left(\max_{k \neq l} A^\mathcal{Q}_{kl},A^\mathcal{Q}_{ll}+1 \right)}\quad l=1,...,P
\end{equation}

\section{Implementation}\label{Implementation}





\begin{figure}[h]
    \centering
    \includegraphics[height=5in]{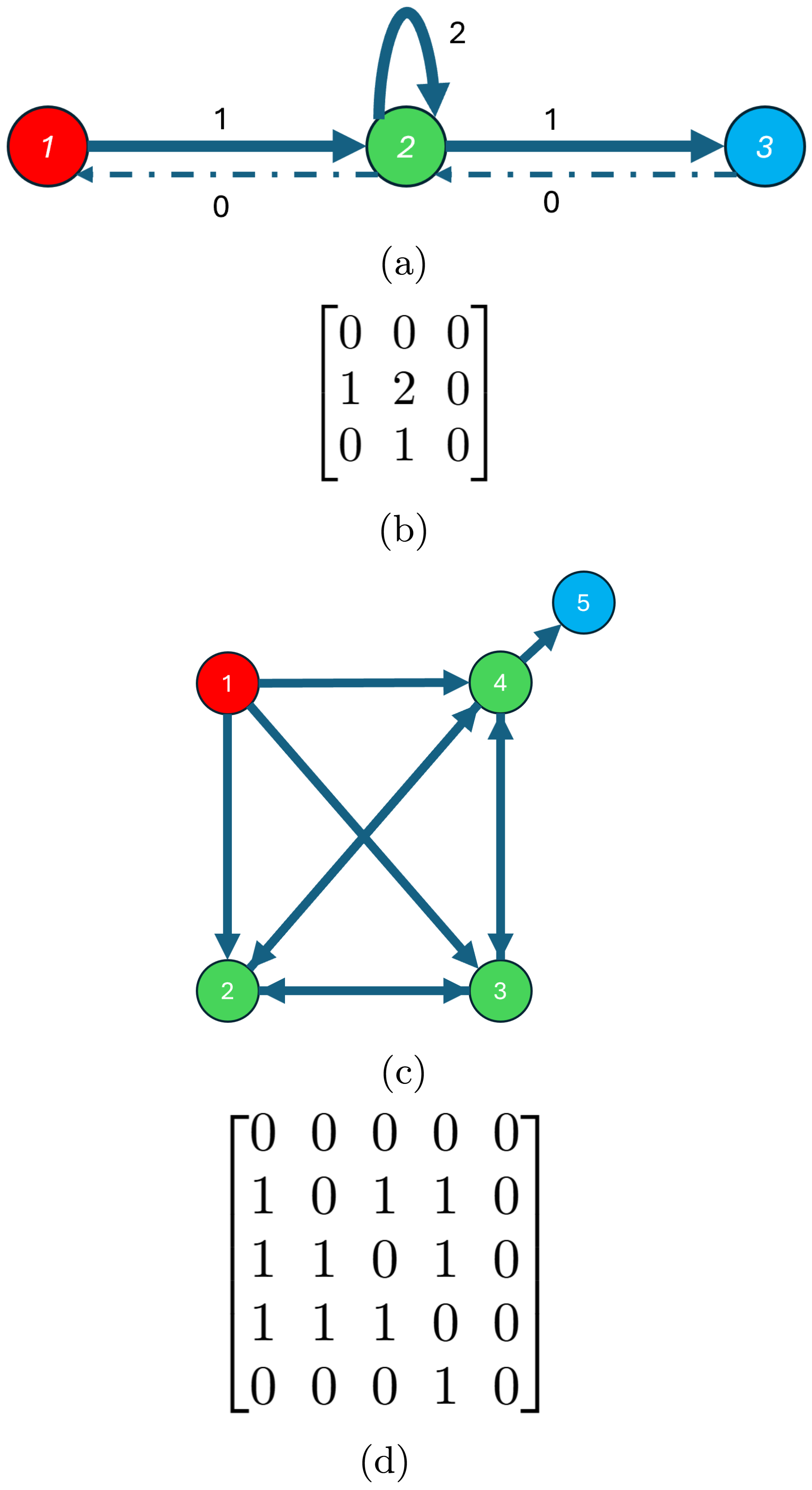}
\caption{Example network for the quotient network expansion algorithm: (\ref{Ex. Network Expansion}a) Simple $P=3$-cluster quotient network example. Some zero-weight edges are added for clarity as dashed lines. (\ref{Ex. Network Expansion}b) The adjacency matrix of the quotient network contained in panel \ref{Ex. Network Expansion}a}. (\ref{Ex. Network Expansion}c) One of the three possible minimal network expansions of the network in panel \ref{Ex. Network Expansion}a. Cluster two in the quotient network highlighted in green becomes three green nodes in the expanded network and so on for the other clusters. (\ref{Ex. Network Expansion}d) The matrix $A$ of the minimal network expansion shown in panel \ref{Ex. Network Expansion}c.\label{Ex. Network Expansion}
\end{figure}

In what follows we consider two different implementations, which we refer to as Case I and Case II. 
In both cases, we start from knowledge of a quotient network, with assigned adjacency matrix $A^{Q}$.
In Case I we attempt to produce a random minimal expansion of the given quotient network, which we do by generating the integer $n_l,l=1,...,P$ according to Eq.\ \eqref{minimimal}. In Case II we attempt to produce a random expansion of a given quotient network with a user-specified number of nodes in each cluster, $n_l,l=1,...,P$. For this case, the integer tuple $n_1, n_2,..., n_P$ is provided and a corresponding expanded network is generated if the user-provided expansion is feasible, i.e., if constraints \eqref{Feasibility Condition Equations} are satisfied.

Let \(\mathcal{R=(V^R,E^R,}A^\mathcal{R}\mathcal{)}\) be the specific quotient network that we are using as input and \(\mathcal{H=(V,E,}A\mathcal{)}\) be the expanded network that can be generated from $\mathcal{R}$, provided that the expansion of \(\mathcal{R}\) is feasible. The index of clusters in \(\mathcal{R}\) is given by the predefined set \(\mathcal{V^R}\). Our method involves the introduction of specific sets that we will use to generate the expansion. Namely, the integer sets $n=\{n_l\}\text{ and }s=\{s_l\},\text{ }l=1,...,P$,  provide the number of nodes per cluster and the number of self loops per cluster, respectively. The entries of $s$ are such that $s_l=(A^\mathcal{R}_{ll})$.
We then consider the sets of integer pairs, $c$ and $f$. 
Specifically, each element of the set $c$ is a pair of clusters $(l,k),l\neq k$ that are connected, such that either cluster $l$ has directed edges to cluster $k$ or vice versa or they are coupled in both directions. The corresponding element of the set $f$ is $(A_{kl}^\mathcal{R},A_{lk}^\mathcal{R})$, i.e., the number of directed connections from cluster $l$ to cluster $k$ and from cluster $k$ to cluster $l$, respectively. 

 Finally, we explain how to construct the adjacency matrix $A$ of the expanded network $\mathcal{H}$. This matrix has $n_1+n_2+...+n_P$ rows and columns. Without loss of generality, we can assign nodes $1,...,n_1$ to be in cluster $1$, nodes $n_1+1,...,n_1+n2$ to be in cluster $2$, and so on. For every cluster pair in $c,f$, with $k$ being the child cluster and $l$ being the parent cluster, we need to ensure that each node in cluster $k$ receives $A_{lk}^{\mathcal{R}}$ directed edges from the nodes in cluster $l$. That can be done
 by adding $A_{lk}^{\mathcal{R}}$ connections from randomly chosen edges in the parent cluster $l$ to each one of the nodes in the child cluster $k$ (with no repetitions.)
 Also, for each
  cluster $l$ and for each node in that cluster, we need to ensure that it receives $A_{ll}^{\mathcal{R}}$ directed edges from other nodes of that cluster (with no repetitions.) That can be done by randomly selecting for each child node in cluster $l$
 $A_{ll}^{\mathcal{R}}$ different parent nodes within the same cluster.

Figure \ref{Ex. Network Expansion} provides an example of a minimal expansion of the quotient network \(\mathcal{R=(V^R,E^R,}A^\mathcal{R}\mathcal{)}\) represented as a graph and adjacency matrix in panels \ref{Ex. Network Expansion}a and \ref{Ex. Network Expansion}b, respectively. We use the generalized example in Fig.\ \ref{Ex. Network Expansion}  to describe the algorithm when Case I applies. By imposing the network feasibility condition of Eq.\ \eqref{Feasibility Condition Equations}, we obtain the number of nodes per cluster and assign them to \(n\). We obtain $n=\begin{bmatrix}1&3&1\end{bmatrix}$, i.e. $n_1=1$ node in cluster $\mathcal{C}_1$, $n_2=3$ nodes in cluster $\mathcal{C}_2$ and $n_3=1$ node in cluster $\mathcal{C}_3$.  Then, in order to generate the set \(s\), we inspect the entries on the main diagonal of $A^\mathcal{R}$ and assign $s=\begin{bmatrix}0 &2&0\end{bmatrix}$, i.e. $A^\mathcal{R}_{11}=0$ self loops in $\mathcal{C}_1 $, $A^\mathcal{R}_{22}=2$ self loops for $\mathcal{C}_2$, and $A^\mathcal{R}_{33}=0$ self loops for $\mathcal{C}_3$.  
Lastly, by imposing Eq.\ \eqref{Outgoing Edge Condition}, we generate the sets  $c=\begin{bmatrix}1,2&2,3\end{bmatrix}$ and $f=\begin{bmatrix}1,0&1,0\end{bmatrix}$, indicating that each node in $\mathcal{C}_2$ will receive $A^\mathcal{R}_{21}=1$ edges from nodes of $\mathcal{C}_1$ (none going in the other direction) and each node in $\mathcal{C}_3$ will receive $A^\mathcal{R}_{21}=1$ edges from nodes of $\mathcal{C}_2$ (none going in the other direction.)


When the algorithm completes its process, the code will plot the graph of \(A\) as seen in panel \ref{Ex. Network Expansion}c (one of the three possible outcomes under Case I), with colored partitions and save the node/edge lists as tables to separate ".csv" file extensions.  The adjacency matrix $A$ is shown in panel \ref{Ex. Network Expansion}d.



\section{Results}\label{Results}

Here we use our algorithm to generate expansions of given quotient networks, both minimal and non-minimal. The clusters of these networks are represented as balanced coloring partitions, in which the nodes in each cluster are given the same color. We decide whether we want an output under Case I or Case II.

\begin{figure}[h!]
    \centering
    \includegraphics[height=5in]{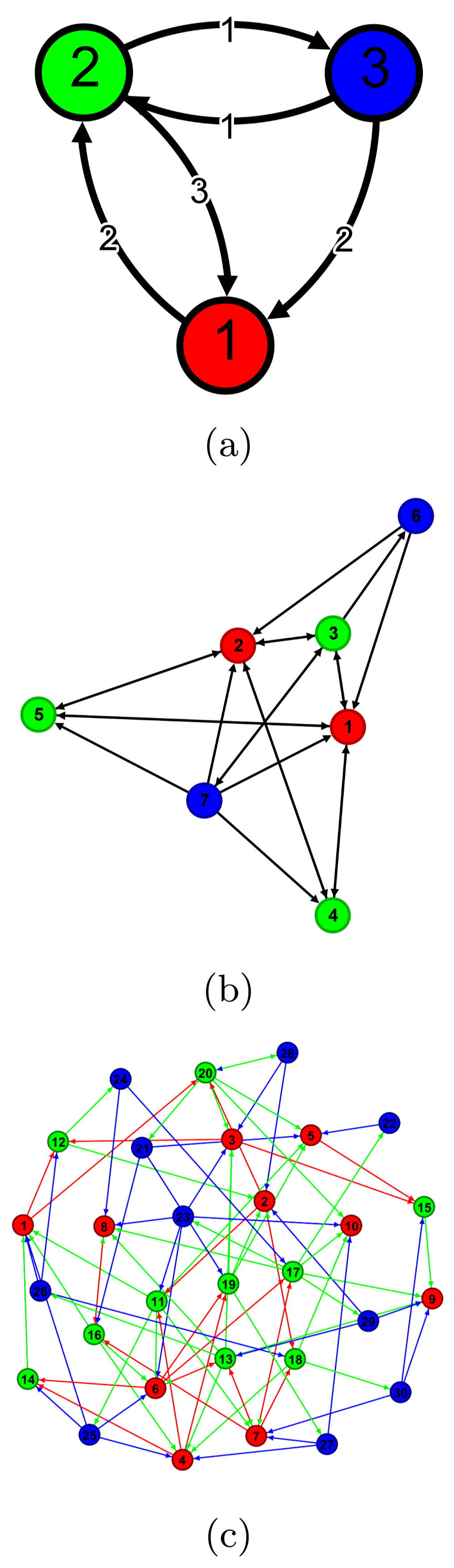}
        \caption{Case I and II example of network expansion utilizing input $\mathcal{R}$: (\ref{Minimal Expansion Example}a) A basic quotient network example in which $\mathcal{R}$ is composed of the values given in the corresponding paragraph of Section \ref{Results}. Cluster one is highlighted in red, then expanded into two red nodes as shown in panel \ref{Minimal Expansion Example}b. The same style follows for the remaining two clusters. (\ref{Minimal Expansion Example}b) A random minimal network expansion generated from $\mathcal{R}$ using the proposed algorithm. The expanded network is generated under Case I, leading to one of 72 possible minimal network expansions. (\ref{Minimal Expansion Example}c) A Case II example of random network expansion of $\mathcal{R}$. Given a user-set number of nodes per cluster where $n=10,10,10$ we generate a 30-node expanded network. The color of each edge matches that of its source node, e.g., an edge generated by a node in red cluster will also be colored red.}
        \label{Minimal Expansion Example}
\end{figure}

Figure \ref{Minimal Expansion Example} is a Case I and II example of the minimal expansion of a quotient network consisting of $P=3$ clusters, resulting in a random output of either a minimal or non-minimal expansion, respectively. The quotient network $\mathcal{R}$ for Fig.\ \ref{Minimal Expansion Example} is defined in terms of \[\mathcal{V^R}=\{1,2,3\}\]\[\mathcal{E^R}=\{(2,1),(1,2),(3,2),(2,3),(1,3)\}\]\[A^\mathcal{R}=\begin{bmatrix}0&3&2\\2&0&1\\0&1&0\end{bmatrix}\]and, when satisfying conditions \eqref{Feasibility Condition Equations}, it is then computed to yield the following sets under Case I: 
$n=\begin{bmatrix}2&3&2\end{bmatrix}$,
$s=\begin{bmatrix}0&0&0\end{bmatrix}$,
$c=\begin{bmatrix}2,1&3,2&1,3\end{bmatrix}$,
$f=\begin{bmatrix}3,2&1,1&0,2\end{bmatrix}$.
We see that there are ${n}_1=2$ nodes for cluster one and $n_2=3$ nodes for cluster two and $n_3=2$ for cluster three. The integer set $s$ indicates there are no self loops in $\mathcal{R}$. From the sets $c,f$ we see that there are connections between all cluster pairs, with the only exception of directed edges from $\mathcal{C}_1$ to $\mathcal{C}_3$, as reflected in panel \ref{Minimal Expansion Example}a between the red and blue clusters. Finally, the sets are processed to generate the adjacency matrix $A$ represented by the graph of $\mathcal{H}$ shown in panel \ref{Minimal Expansion Example}b. Under Case II, the same process occurs, but  with a set $n=\begin{bmatrix}10&10&10\end{bmatrix}$ that we have specified. We have checked that this expansion is feasible. A new random expanded network adjacency matrix $A$ is produced under Case II, which is represented by the graph in panel \ref{Minimal Expansion Example}c.

\begin{figure}[h!]
    \centering
    \includegraphics[width=1\linewidth]{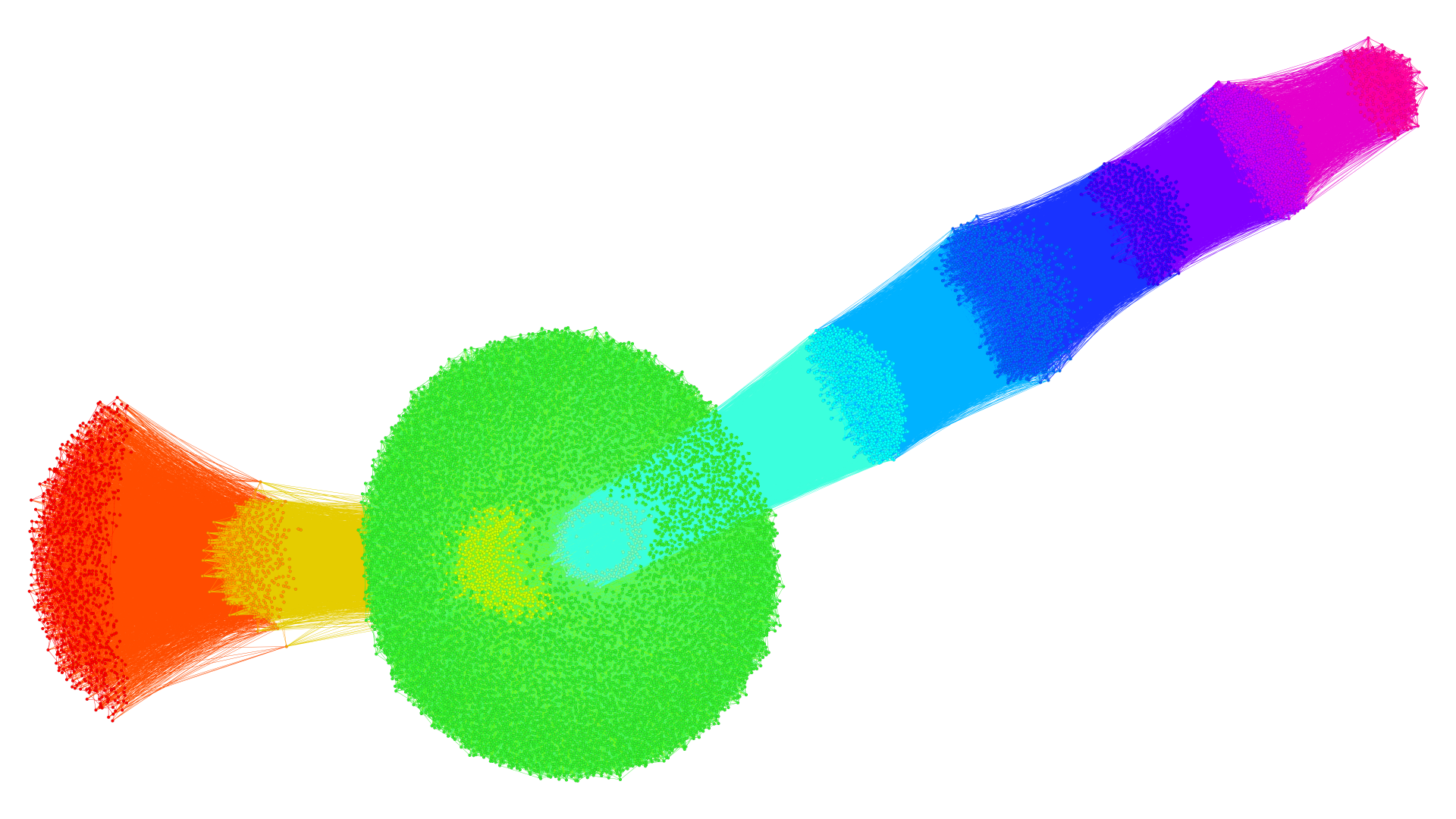}
    \caption{A Case II $P=10$-cluster high-load network example of over 10,000 nodes, generated in under 20 seconds. The color of each edge matches that of its source node,}
    \label{High-Load Network}
\end{figure}

Figure \ref{High-Load Network} is a Case II high node-count random expanded network consisting of more than 10,000 nodes generated from an arbitrary $P=10$-cluster quotient network $\mathcal{R}$ in a total process time of less than 20 seconds. We set the number of nodes per cluster to be $n=1000, 400,500,10000,200,800,950,500,600,200$. We first confirm that $\mathcal{R},n$ satisfy constraints \eqref{Feasibility Condition Equations}; we then compute the sets $s,c,f$, which we use to produce the large expanded network shown in Fig.\ \ref{High-Load Network}.

\begin{figure}[h!]
    
\end{figure}

\section{Conclusions}

Real complex networks are directed and present symmetries. It is therefore essential to be able to generate synthetic networks that present both these features. 
In this paper we have shown how to generate an expanded network from knowledge of its quotient network and a desired number of nodes for each one of the clusters. First, we address the question whether such an expansion is feasible. Then, in case the answer is yes, we
present an algorithm that enables a user to generate an expanded network from two inputs: a quotient network and a desired number of nodes for each cluster of the expanded network. {As the algorithm specifies the number of nodes in each cluster of the expanded network, it produces a network with a given degree sequence, similar to the configuration model
 \cite{Mo:Re95}.}

Our work differs from \cite{klickstein2018generating,klickstein2018generating2}, which were the first papers to study generating algorithms for networks with symmetries, {as our algorithm produces expanded networks that are directed. }
Directed networks exhibit significantly more flexibility for expansion than undirected networks. In many cases, undirected network expansions are not feasible, depending on the integer relation of edges between nodes and how many nodes are expected in each cluster \cite{klickstein2018generating}. Conversely, the feasibility of directed networks depends only on the number of nodes desired for each cluster, where the number of nodes in the minimal expansion, see Eq.\ \eqref{minimimal}, provides a lower threshold. {A limitation of this work is that it focuses on unweighted directed graphs, while many real networks are weighted, with possibly non-integer `noisy' weights. The generation of weighted directed graphs would require an alternative definition of `cluster' \cite{nathe2022looking} and is beyond the scope of this paper.}

\section*{ACKNOWLEDGMENTS}
We acknowledge support from grants AFOSR FA9550-24-1-0214 and Oak Ridge National Laboratory 006321-00001A.

\section*{AUTHOR DECLARATIONS}

\textbf{Conflict of Interest.}
The authors have no conflicts to disclose.

\section*{DATA AVAILABILITY}
The data that support the findings of this study are available
within the article.

\section*{CODE AVAILABILITY}
{All computational procedures described in this work are fully reproducible;} we provide an open-source code that is available for download at this link \url{https://github.com/Joma101/Quotient_Network_Expansion_Algorithm}. 

\bibliographystyle{plain}

\newcommand{\noop}[1]{}


\end{document}